\documentclass[11pt]{article}
\usepackage{epsfig,graphics,wrapfig,moriond}

\def\be{\begin{equation}}
\def\ee{\end{equation}}
\def\bea{\begin{eqnarray}}
\def\eea{\end{eqnarray}}

\def\psidp{\psi^{\prime\prime}}
\def\ipb{\rm pb^{-1}}
\def\bfr{ {\cal{B}} }
\def\ell{\cal{l}}

\begin{document}
\vspace*{4cm}

\title{Measurements of Hadronic, Semileptonic and Leptonic Decays
of $D$ Mesons at $E_{cm}$=3.77 GeV in CLEO-c}

\author{ Steven R. Blusk }

\address{Syracuse University, Department of Physics, 201 Physics Building,\\
Syracuse, NY USA
}

\maketitle\abstracts{
	We report on recent measurements of hadronic, semileptonic and leptonic
$D$ decays taken using $\sim$57 $\ipb$ of data collected on the $\psidp$ resonance 
using the CLEO-c detector. 
}

\section{Introduction}

	Study of $D$ meson decays provides a laboratory in which to study
both the strong and weak interactions. On the $\psidp$ resonance, $D\bar{D}$
pairs are produced in a pure $J^{PC}=1^{--}$ state with no additional particles,
leading to a powerful kinematic constraint: $E_D=E_{beam}$. Combining
this constraint with the excellent hermiticity and particle identification 
of the CLEO-c detector~\cite{cleoc}, allows for clean measurements in $D$ decays which
have traditionally been statistically limited, difficult or inaccessible.

	The analyses presented here include about 57 $\ipb$ of data collected on 
the $\psidp$ resonance using the CLEO-c detector. They all proceed by fully 
reconstructing one $D$ meson in an event (hereafter, referred to a $D_{\rm tag}$), 
and studying the decay of the second ($D_{\rm sig}$). Tagged samples with
excellent signal-to-background (S:B) ratios are achieved by requiring that the energy of
the $D$ candidate is consistent with the beam energy. One then can then
obtain a mass resolution of order 2 MeV by replacing the $D$ energy with the
beam energy, resulting in S:B ratios in Cabibbo-favored modes
ranging from $\sim$200:1 to $\sim$5:1. Throughout this paper, charge conjugate 
modes are implied.


\vspace{-0.15in}
\section{Hadronic D Decays}

	Precise measurement of hadronic $D$ decays are of great importance
since, for example, they are often used to normalize $B$ and $D$ semileptonic
branching fractions, which in turn are used to extract CKM matrix 
elements~\cite{ckm}. The hadronic analysis reconstructs single-tagged and 
double-tagged events, where one or both $D$ mesons decay into 
$D^0\to K^-\pi^+,~K^-\pi^+\pi^0,~K^-\pi^+\pi^+\pi^-$ 
or $D^+\to K^-\pi^+\pi^+$, $K^-\pi^+\pi^+\pi^0$, $K^0_s\pi^+$, $K^0_s\pi^+\pi^0$,
$K^0_s\pi^+\pi^+\pi^-$, $K^+K^-\pi^+$.
It is straightforward to show that the individual branching fractions
are given by ${\cal{B}}_i={N_{ij}\over N_j}{\epsilon_j\over \epsilon_{ij}}$
and the number of $D\bar{D}$ pairs by 
$N_{D\bar{D}}={N_i N_j\over 2N_{ij}}{\epsilon_{ij}\over \epsilon_i\epsilon_j}$,
where $N_{ij}$ is the yield of double-tagged events with tag mode $j$
and signal mode $i$, $N_j$ is the yield of single-tagged events in mode
$j$, and $\epsilon_{ij}$, $\epsilon_{j}$ are their respective 
reconstruction efficiencies. As $\epsilon_{ij}\approx\epsilon_i\epsilon_j$,
the measured ${\cal{B}}_i$'s are nearly independent of the tag modes' efficiencies 
and almost all reconstruction systematics cancel in $N_{D\bar{D}}$. 
The 9 ${\cal{B}}_i$'s, $N_{D^0\bar{D^0}}$, and $N_{D^+D^-}$ are extracted
through a $\chi^2$ minimization involving the difference between
the expected background-subtracted, efficiency-corrected single and 
double-tagged yields and the corresponding observed yields. The 
average single tag yields, their efficiencies and the corresponding
branching fractions are shown in Table~\ref{tab:hadbf}. The fitted number
of $D^0\bar{D^0}$ and $D^+D^-$ pairs are $(2.06\pm0.038\pm0.02)\times 10^5$ and
$(1.56\pm0.04\pm0.01)\times 10^5$, where the indicated uncertainties are
statistical and systematic. These yields correspond to peak cross-sections
$\sigma(e^+e^-\to D^0\bar{D^0})=3.60\pm0.07^{+0.07}_{-0.05}$ nb
and $\sigma(e^+e^-\to D^+D^-)=2.79\pm0.07^{+0.10}_{-0.04}$ nb. The 
systematic uncertainties include particle reconstruction (0.7\%
per charged particle, 3\%/$K^0_s$, 2\%/$\pi^0$), 
particle identification (0.3\%/$\pi^{\pm}$, 0.7\%/K$^{\pm}$),
event selection (1.5\%), $\psidp$ natural width (0.6\%), 
final state radiation (0.5\% [1.0\%] for single [double] tags),
resonant substructure (0.4-1.5\%), doubly-Cabibbo suppressed 
interference (0.8\%), and signal fitting (0.5\%). Further details of this 
analysis can be found in the references~\cite{dhadronic}.

\begin{table}[ht]
\caption{Summary of branching fraction measurements showing the
average number of reconstructed single tags, the
average reconstruction efficiency, and the branching fraction 
measurements in $D$ decays.
\label{tab:hadbf}}
\vspace{-0.05cm}
\begin{center}
\begin{tabular}{|l|c|c|c|c|}
\hline
Mode  		  &  $<N_D>$  ($\times 10^3$)  &  $<\epsilon_D>$ (\%) & $\bfr$  (\%) & $\bfr$(\%) PDG~\cite{pdg} \\
\hline
$D^0\to~K^-\pi^+$ 	  & 5.13$\pm$0.07 &  65.7$\pm$0.3 & $3.91\pm0.08\pm0.09$ & $3.80\pm0.09$ \\
$D^0\to~K^-\pi^+\pi^0$   & 9.49$\pm$0.07 &  33.2$\pm$0.1  & $14.94\pm0.30\pm0.47$  & $13.0\pm0.8$ \\
$D^0\to~K^-\pi^+\pi^+\pi^-$  & 7.44$\pm$0.07 &  44.6$\pm$0.2 &  $8.29\pm0.17\pm0.32$ & $7.46\pm0.31$ \\
\hline
$D^+\to~K^-\pi^+\pi^+$ 	    & 7.56$\pm$0.09 &  51.7$\pm$0.2 & $9.52\pm0.25\pm0.27$ & $9.2\pm0.6$ \\
$D^+\to~K^-\pi^+\pi^+\pi^0$ & 2.42$\pm$0.07 &  27.2$\pm$0.2 & $6.04\pm0.18\pm0.22$ & $6.5\pm1.1$ \\
$D^+\to~K^0_s\pi^+$ 	    & 1.12$\pm$0.04 &  45.6$\pm$0.4 & $1.55\pm0.05\pm0.06$ & $1.41\pm0.10$ \\
$D^+\to~K^0_s\pi^+\pi^0$  & 2.55$\pm$0.07 &  23.4$\pm$0.2     & $7.17\pm0.21\pm0.38$ & $4.9\pm1.5$ \\
$D^+\to~K^0_s\pi^+\pi^+\pi^-$ & 1.61$\pm$0.06 &  31.4$\pm$0.2 & $3.20\pm0.11\pm0.16$  & $3.6\pm0.5$\\
$D^+\to~K^+K^-\pi^+$ 	    & 0.63$\pm$0.03 &  42.6$\pm$0.5 &  $0.97\pm0.04\pm0.04$ & $0.89\pm0.08$\\
\hline
\end{tabular}
\vspace{-0.25in}
\end{center}
\end{table}

\section{Semileptonic Decays}

	Semileptonic widths for $D\to X_{s(d)}l^+\nu$ directly probe the elements
of the CKM matrix. When $J^{P}(X_{(s,d)})=0^-$, the differential width 
is given by:

\vspace{-0.12in}
\begin{equation}
{d\Gamma(D\to X_{s(d)}l^+\nu)\over dq^2} = {|V_{cs(cd)}|^2 p^3\over 24\pi^3} f(q^2),
\end{equation}

\vspace{-0.05truein}
\noindent where $q^2$ is the momentum transfer squared between the $D$ meson and the 
final state hadron in the $D$ rest frame. While the normalization of the form factor, 
$f(q^2)$, 
must be obtained from theory (lattice QCD, for example), CLEO-c can measure its 
shape and in doing so provide a stringent test of theoretical predictions. 
Provided theory is able to describe the shape of $f(q^2)$, we then use this
tested theory for the normalization, which then permits immediate extraction 
of $V_{cs}$ and $V_{cd}$. 

	The semileptonic analyses use 9 $D^0_{\rm tag}$ and 6 $D^+_{\rm tag}$ modes.
For each tagged event, we search for an electron with $p>$200 MeV/$c$ and
additional hadrons recoiling against $D_{\rm tag}$. Charged kaons and pions
are separated using a combined $dE/dx$ and RICH information, with an 
efficiency of about 95\% and a fake rate of no more than a few percent. 
Candidate $\pi^0$'s are formed by pairing two photons and requiring 
$|M_{\gamma\gamma}-M_{\pi^0}|<3\sigma$. $K^0_s$ candidates are defined
as $\pi^+\pi^-$ pairs which have an invariant mass within 12 MeV/$c^2$ 
of the $K^0_s$ mass. Similarly, $\rho^0$ ($\rho^-)$ candidates are 
formed by pairing $\pi^+\pi^-$ ($\pi^-\pi^0$) and requiring 
$|M_{\pi\pi}-M_{\rho}|<150$ MeV/$c^2$. Finally, we reconstruct 
$\omega$ candidates using $\pi^+\pi^-\pi^0$ combinations with
$|M_{\pi^+\pi^-\pi^0}-M_{\omega}|<20$ MeV/$c^2$.

\begin{figure}
\begin{minipage}[t]{3.0in}
\includegraphics[width=0.9\textwidth]{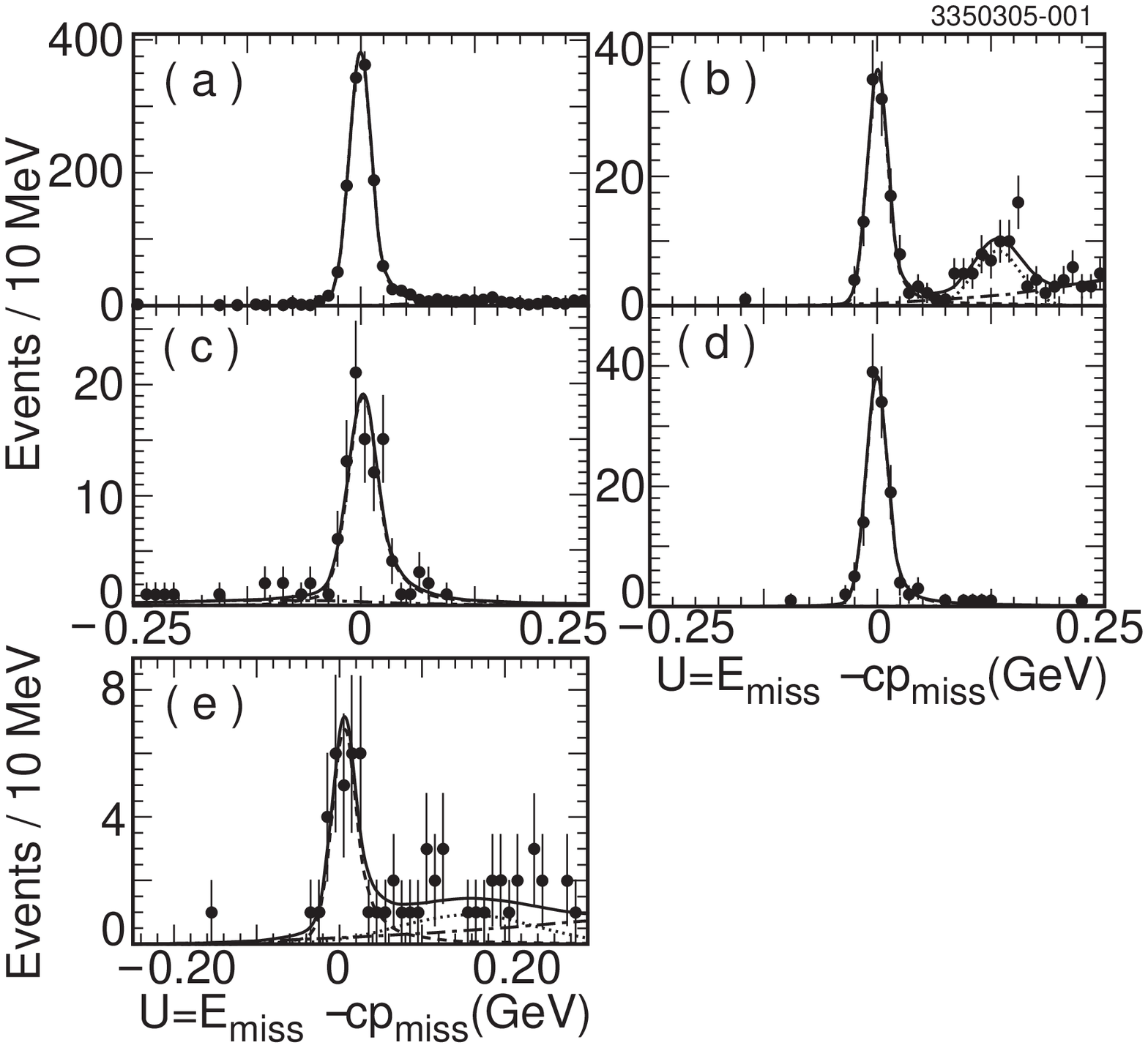}
\vspace{-0.15in}
\caption{Distributions in $U=E_{miss}-|\vec{p}_{miss}|$ for (a) $D^+\to K^0_s e^+\nu_e$,
(b)  $D^+\to K^{*0} e^+\nu_e$, (c)  $D^+\to\pi^0 e^+\nu_e$, (d)  $D^+\to\rho^0 e^+\nu_e$, 
and (e)  $D^+\to\omega e^+\nu_e$.}
\label{fig:dch_sl}
\end{minipage}
\hfill
\begin{minipage}[t]{3.0in}
\includegraphics[width=0.9\textwidth]{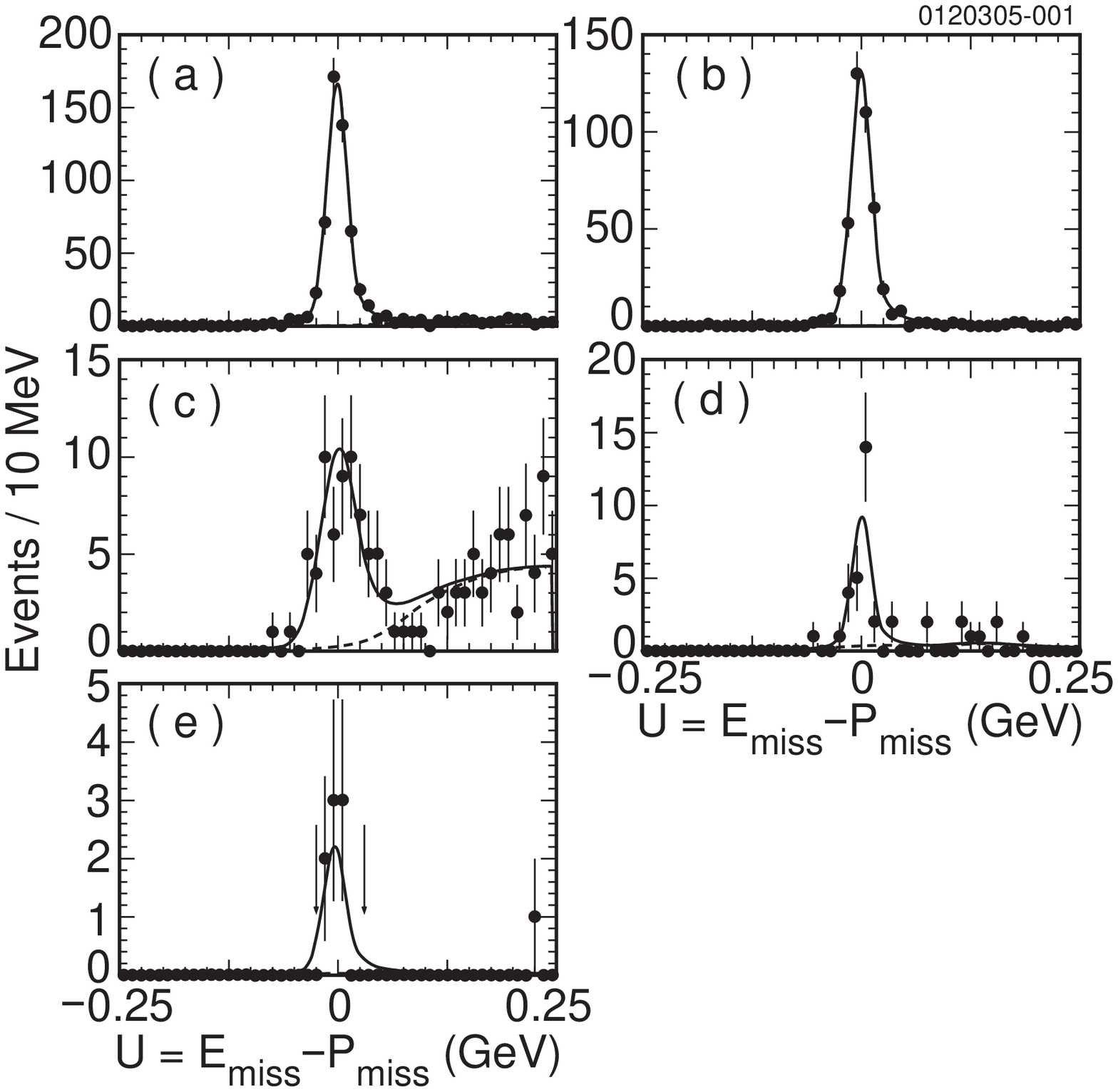}
\vspace{-0.1in}
\caption{Disributions in $U=E_{miss}-|\vec{p}_{miss}|$ for (a) $D^0\to K^- e^+\nu_e$,
(b)  $D^0\to \pi^- e^+\nu_e$, (c)  $D^0\to K^{*-}[K^-\pi^0] e^+\nu_e$, 
(d)   $D^0\to K^{*-}[K^0_s\pi^-] e^+\nu_e$, and 
(e)  $D^+\to\rho^- e^+\nu_e$.\label{fig:d0_sl}}
\end{minipage}
\hfill
\vspace{-0.25in}
\end{figure}

	For each event, we require that there be no additional charged tracks
beyond the $D_{\rm tag}$ and the semileptonic candidate. 
Properly reconstructed events are identified by a peak at zero in the quantity 
$U=E_{miss}-|\vec{p}_{miss}|$. Here, $E_{miss}$ and $\vec{p}_{miss}$
are the missing energy and missing momentum in the semileptonic decay, both of
which are calculable on an event-by-event basis. The $U$ distributions for 
$D^+\to (K^0_S, \bar{K^{*0}}[K^-\pi^+], \pi^0, \rho^0, \omega) e^+\nu_e$ are shown in 
Figs.~\ref{fig:dch_sl}(a)-(e) respectively. The analogous distributions for
$D^0\to (K^-, \pi^-, K^{*-}[K^-\pi^0], K^{*-}[K^0_s\pi^-], \rho^-) e^+\nu_e$
are shown in Figs.~\ref{fig:d0_sl}(a)-(e), respectively. The $D^+\to\omega e^+\nu_e$
and $D^0\to\rho^- e^+\nu_e$ constitute first observations of these decays.
The yields, efficiencies and corresponding branching fractions are shown
in Table~\ref{tab:slbf}. These measurements are already better than world
average, and substantially more data are imminent. These analyses will soon
be submitted for publication.

\begin{table}[t]
\caption{Summary of semileptonic branching fraction measurements showing the
average reconstruction efficiency, number of reconstructed events, reconstruction 
efficiency and the branching fractions.
\label{tab:slbf}}
\vspace{-0.05cm}
\begin{center}
\begin{tabular}{|l|c|c|c|c|}
\hline
Decay  		  &  Yield  &  $\epsilon_D$ (\%) & $\bfr$  & $\bfr$(\%) PDG~\cite{pdg} \\
\hline
$D^0\to K^- e^+\nu_e$  & 1311.0$\pm$36.6 &  63.6$\pm$0.5  & $(3.44\pm0.10\pm0.10)\%$  & $3.58\pm0.18$ \\
$D^0\to\pi^- e^+\nu_e$  & 116.8$\pm$11.2 &  74.2$\pm$0.5  & $(2.62\pm0.25\pm0.08)\times10^{-3}$ & $0.36\pm0.06$ \\
$D^0\to K^{*-}[K^-\pi^0] e^+\nu_e$  & 94.1$\pm$10.4 &  22.0$\pm$0.3  & $(2.16\pm0.24\pm0.12)$ &  \\
$D^0\to K^{*-}[K^0_s\pi^-] e^+\nu_e$  & 125.2$\pm$11.6 &  40.4$\pm$0.4  & $(2.25\pm0.21\pm0.13)\%$  & $3.58\pm0.18$ \\
$D^0\to K^{*-}e^+\nu_e$  &            &               & $(2.21\pm0.16\pm0.09)\%$ &  $2.15\pm0.35$ \\
$D^0\to\rho^- e^+\nu_e$  & 31.1$\pm$3.3 &  27.0$\pm$0.4  & $(1.94\pm0.39\pm0.13)\times10^{-3}$ &  \\
\hline\hline
$D^+\to K^0_s e^+\nu_e$  & 545$\pm$24 &  57.1$\pm$0.4  & $(8.71\pm0.38\pm0.37)\%$   & $6.7\pm0.9$ \\
$D^+\to K^{*0} e^+\nu_e$  & 422$\pm$21 &  34.8$\pm$0.3  & $(5.70\pm0.28\pm0.25)\%$   & $5.5\pm0.7$ \\
$D^+\to\pi^0 e^+\nu_e$  & 63$\pm$9 &  45.2$\pm$1.0  & $(0.44\pm0.06\pm0.03)\%$   & $0.31\pm0.15$ \\
$D^+\to\rho^0 e^+\nu_e$  & 27$\pm$6 &  40.0$\pm$1.1  & $(0.21\pm0.04\pm0.02)\%$   & $0.25\pm0.10$ \\
$D^+\to\omega e^+\nu_e$  & 8.0$\pm$2.8 &  16.4$\pm$0.6  & $(0.17\pm0.06\pm0.01)\%$  & \\
\hline
\end{tabular}
\end{center}
\end{table}


	
\vspace{-0.05in}
\section{Leptonic Decays}
\vspace{-0.05in}

	Leptonic decays of $D$ mesons provide direct access to the decay constant $f_D$, which
is related to the wave-function overlap probability for the constituent $c$ and $\bar{d}$ quarks.
An accurate measurement of $f_{D^+}$ provides a stringent test of lattice QCD and other QCD-inspired models
to predict decay constants, and in particular, $f_B$ and $f_{B_s}$.
	
\begin{wrapfigure}{r}{6cm}
\begin{center}
\vspace{-0.1in}
\scalebox{0.35}
{
\includegraphics{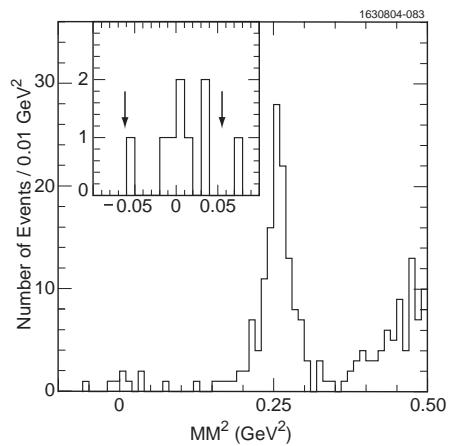}
}
\end{center}
\vspace{-0.15in}
\caption{Disributions in $MM^2$ using $D^-$ tags and one recoiling charged track. 
}
\label{fig:dlep}
\end{wrapfigure}

	From a single-tag sample consisting of tagged $D^-$ decays in $K^+\pi^-\pi^-$, 
$K^+\pi^-\pi^-\pi^0$, $K^0_s\pi^-$, $K^0_s\pi^-\pi^0$, and $K^0_s\pi^+\pi^-\pi^-$, we
select a subset of events which contain exactly one extra charged particle. The
muon system is not used in this analysis, since most muons in $D^+\to\mu^+\nu_{\mu}$ are
below its momentum threshold of about 1 GeV.  We also require that there is no
additional shower with energy exceeding 250 MeV. Signal
$D^+\to\mu^+\nu_{\mu}$ events are distinguishable by a peak at zero in the missing-mass
($MM$) spectrum, where $MM^2=(E_{\rm beam}-E_{\mu})^2-(-\vec{p_{D^-}}-\vec{p_{\mu^+}})^2$
(=$m_\nu^2$, for $D^+\to\mu^+\nu_{\mu}$). The distribution in $MM^2$ for these events is
shown in Fig.~\ref{fig:dlep}. The large peak at 0.25 GeV$^2$ is from $D^+\to K^0_L\pi^+$.
The inset shows a blowup of the region near zero, and shows the 8 signal candidates which 
lie within the signal region, defined to be $\pm$2$\sigma$.

The primary sources of background come from $D^+\to\pi^+\pi^0$ (0.31$\pm$0.04 events), 
where the $\pi^0$ is not detected, and $D^+\to\tau^+\nu_{\tau}$, followed by 
$\tau^+\to\pi^+\nu_{\tau}$ (0.30$\pm$0.07 events). Other backgrounds
include $D^+\to K^0_L\pi^+$ (0.06$\pm$0.05 events), $D^0\bar{D^0}$ (0.17$\pm$0.17 events)
and continuum (0.16$\pm$0.16 events). In total, the expected background is 1.00$\pm$0.25 
events.

	The $D^+\to\mu^+\nu_{\mu}$ branching fraction is readily computed
using $N_{\rm tag}$=28574$\pm$629, $N_{D\to\mu^+\nu_{\mu}}=7.0\pm2.8$ and a 
single-muon detection efficiency of (69.9$\pm$3.7)\%, from which we find
$\bfr = (3.5\pm1.4\pm1.6)\times 10^{-4}$. Taking $\tau_{D^+}=1.04$ ps, and
$V_{cd}=0.224$~\cite{pdg}, we find $f_{D^+}=(202\pm41\pm17)$ MeV. This is the
most precise measurement of this quantity and compares well with lattice QCD
and a number of other models. As more data are collected, we will be in a position
to discriminate among these predictions. More details can be found in 
in the references~\cite{dleptonic}.

\vspace{-0.15in}
\section*{Acknowledgments}
	I would like to thank my CLEO colleagues for their input on these
analyses. This work was supported by the National Science Foundation and the
U.S. Dept. of Energy.


\section*{References}
\begin{thebibliography}{99}

\bibitem{cleoc}
The CLEO-c detector is an upgrade from the CLEO-III detector in which the silicon
detector was replaced with a six-layer vertex drift chamber. For details on the
CLEO detector, refer to:
CLEO Collaboration, Y. Kubota {\it et al.,} Nucl. Instrum. Meth. Phys. Res. {\bf A320}, 66 (1992); G. Viehauser {\it et al.,} Nucl. Instrum. Meth. A{\bf 462}, 146 (2001); 
D. Peterson {\it et al.,} Nucl. Instrum. Meth. A{\bf 478}, 142 (2002); 
M. Artuso {\it et al.,} Nucl. Instrum. Meth. A{\bf 502}, 91 (2003). 
The CLEO-c yellow book is available as CLEO Document CLNS-01/1742.

\bibitem{ckm} M. Kobayashi and T. Maskawa, Prog. Theor. Phys. {\bf 49}, 652 (1973).


\bibitem{dhadronic} CLEO Collaboration, Q. He {\it et al.,} Submitted to
Phys. Rev. Lett., [hep-ex/0504003].

\bibitem{pdg}
Particle Data Group, S. Eidelman {\it et al.,} Phys. Lett. {\bf B592}, 1 (2004).

\bibitem{dleptonic}
CLEO Collaboration, G. Bonvinci {\it et al.,} Phys. Rev. D {\bf 70}, 112004 (2004).


\end{thebibliography}
\end{document}